\documentclass{article}

\usepackage{arxiv}

\usepackage[utf8]{inputenc}
\usepackage[T1]{fontenc}
\usepackage{microtype}
\usepackage{graphicx}
\usepackage{booktabs}
\usepackage{array}
\usepackage{enumitem}
\usepackage{xcolor}
\usepackage[breakable]{tcolorbox}
\usepackage{tikz}
\usetikzlibrary{positioning,arrows.meta}
\usepackage{float}
\usepackage{caption}
\usepackage[hyperfootnotes=false]{hyperref}
\usepackage{xurl}
\definecolor{linkblue}{RGB}{0,60,120}
\hypersetup{colorlinks=true,linkcolor=linkblue,urlcolor=linkblue,citecolor=linkblue}
\usepackage[hang]{footmisc}
\setlist{itemsep=2pt,topsep=4pt}

\newcommand{\bibentrystart}{\hangindent=1.5em\hangafter=1\noindent}

\title{Near-Term Verification Methods for AI Chip Exports\thanks{This paper was originally published as a report by the Institute for AI Policy and Strategy (IAPS) in August 2026: \url{https://www.iaps.ai/research/near-term-verification-methods-for-ai-chip-exports}.}}
\date{September 2026}
\renewcommand{\shorttitle}{Near-Term Verification Methods for AI Chip Exports}

\author{
  Bruna Avellar\thanks{This work was partially carried out during the author's fellowship at the Institute for AI Policy and Strategy.} \\
  Independent Researcher \\
  \And
  Erich Grunewald \\
  Institute for AI Policy and Strategy \\
}

\renewcommand{\headeright}{}
\renewcommand{\undertitle}{}

\hypersetup{
  pdftitle={Near-Term Verification Methods for AI Chip Exports},
  pdfauthor={Bruna Avellar, Erich Grunewald},
}

\begin{document}

\maketitle

\begin{abstract}
\noindent AI chip export controls can help the United States shape the development of frontier AI, but their effectiveness depends on reliable methods for verifying compliance. This paper examines near-term verification mechanisms (implementable in approximately one year) and groups them into three categories: end-location verification (whether controlled chips remain in authorized locations and/or jurisdictions), end-user verification (whether entities that acquire or access compute are legitimate), and end-use verification (whether computing power is used for prohibited purposes). We discuss how each mechanism can be implemented within the regulatory framework of the U.S. Bureau of Industry and Security (BIS), outlining implementation steps and identifying which actors can perform verification (BIS, exporters, or accredited third-party auditors). Given BIS's resource constraints, the most viable mechanisms rely on private-sector actors working alongside BIS, leverage existing technologies, and scale without requiring large increases in government staffing. These mechanisms could also help monitor future international agreements on AI.
\end{abstract}

\clearpage
\tableofcontents
\clearpage

\section*{Executive Summary}
\phantomsection\label{executive-summary}
\addcontentsline{toc}{section}{Executive Summary}

AI chip export controls are an indispensable tool for the United States to shape the trajectory of frontier AI development, but their effectiveness depends on how well they are enforced. Enforcement, in turn, requires the ability to verify compliance with export controls. This paper serves as an implementation guide for policymakers seeking to strengthen enforcement of export controls, presenting a menu of near-term verification mechanisms---here defined as implementable in approximately one year---each selected for its feasibility, cost-effectiveness, and potential to serve as meaningful components of an effective export control enforcement regime.

We define ``verification'' as the process of obtaining information that determines, with a high degree of confidence, whether export controls have been violated. The mechanisms we describe are divided into three categories:

\begin{itemize}
\item
  \textbf{End-location verification} ensures that controlled chips remain within authorized locations and/or jurisdictions (\hyperref[end-location-verification]{more}).
\item
  \textbf{End-user verification} ensures the legitimacy of entities that acquire or access compute (\hyperref[end-user-verification]{more}).
\item
  \textbf{End-use verification} ensures that computing power is not being used for prohibited purposes (\hyperref[end-use-verification]{more}).
\end{itemize}

Each of these categories is described in a section below, outlining specific verification mechanisms and discussing how they can be implemented within the regulatory framework of the Bureau of Industry and Security (BIS), the U.S. agency that administers and enforces dual-use export controls.

As of July 2026, BIS is facing significant challenges in implementing and enforcing its existing verification methods, largely due to its relatively small budget and outdated technology.\footnote{Roberts, ``\href{https://www.the-substrate.net/p/bis-is-getting-more-fundingheres}{BIS is getting more funding---here\textquotesingle s how to spend it.}''; Roberts, ``\href{https://www.the-substrate.net/p/bis-should-build-a-lean-mean-data}{BIS should build a lean, mean, data-driven enforcement machine.}''} The Trump administration has requested \$450 million for BIS in fiscal year 2027, but even if forthcoming, this would not fully resolve BIS's resource gap.\footnote{U.S. Department of Commerce, ``\href{https://www.commerce.gov/sites/default/files/2026-04/FY2027-BIS-CJ-Submission.pdf}{Fiscal Year 2027 President's Budget Request.}''} Considering these constraints, the most viable verification approaches today rely on private-sector actors working alongside BIS, leverage existing technologies, and scale without requiring large increases in government staffing.

The verification mechanisms described in this paper can each be carried out by one of three actors: BIS itself, through its existing enforcement infrastructure; an exporting company, either voluntarily or as an export condition imposed by BIS; or an independent third-party auditor accredited by BIS, should such an accreditation system be established. An accreditation system for third-party auditors does not currently exist, but it could help scale enforcement, given BIS's limited resources relative to the volume of controlled items in circulation. A separate IAPS report has described this proposal in detail.\footnote{Aarne and Grunewald, ``\href{https://static1.squarespace.com/static/64edf8e7f2b10d716b5ba0e1/t/69bc642c9cbbd038d27a9660/1773954092111/Export+Auditors+as+Market-Powered+Export+Enforcement.pdf}{Export Auditors as Market-Powered Export Enforcement.}''}

Verification mechanisms vary considerably in their cost, intrusiveness, and enforcement value. An effective mechanism can provide significant value even if it is not impossible to evade---it can be useful as long as it is effective enough to make it more costly for malicious actors to evade export controls.

For end-location verification, one option is on-site inspections. These are effective but labor-intensive and costly, making them best suited as a targeted mechanism for high-risk exports rather than a routine practice. Another option is remote video inspections or auditor-directed video walkthroughs, which offer a less resource-intensive alternative to physical inspections but are more susceptible to staging or manipulation---for instance, using AI to create synthetic video feeds---than physical visits. A third option is a delay-based mechanism, which provides near-real-time information about a chip's location by measuring the response delay between chips and trusted landmark servers, leveraging existing chip-level attestation capabilities at relatively low cost. However, this mechanism offers limited visibility during shipping and warehousing, before chips are deployed. These approaches are not mutually exclusive and can be layered, for example using delay-based mechanisms as a first line of detection and triggering video or on-site inspections when anomalies surface.

For end-user verification, enhanced Know-Your-Customer checks---which include automated supply chain risk analysis---can be an effective method to detect concealed ownership structures and prevent restricted entities from acquiring chips through intermediaries. In this approach, exporters\footnote{Throughout this paper, unless otherwise noted, by ``exporter'' we mean ``exporter, re-exporter, or in-country transferor.'' Likewise, by ``export'' we mean ``export, reexport, or in-country transfer''.} would use specialized platforms to review publicly available information such as corporate filings to identify links to entities of concern, like those subject to specific license requirements per the BIS-maintained Entity List. Although this mechanism works well as a preventive screen to verify the identity of purchasers \emph{before} export, it cannot be used as an ongoing monitoring tool.

For end-use verification, cross-checking end-use declarations against publicly available information that attest to a customer's line of business, such as corporate records, can help chip exporters verify that a customer's declared use is consistent with their actual business activities. This mechanism might also become relevant for cloud providers, should cloud access become regulated under the BIS's Export Administration Regulations (EAR), as they could perform similar cross-checks on customers.

These mechanisms are, however, designed to detect violations rather than enforce compliance on their own. Robust enforcement systems should pair verification with follow-up investigations whenever there is reason to suspect a violation has occurred, and proceed with enforcement action if a violation is confirmed.

Beyond their use for export enforcement, these verification mechanisms could also serve to monitor compliance with future international AI agreements, which may become necessary to coordinate safety measures across states as advanced AI systems grow increasingly capable and pose growing systemic risks.\footnote{Baker et al., ``\href{https://arxiv.org/abs/2507.15916}{Verifying International Agreements on AI: Six Layers of Verification for Rules on Large-Scale AI Development and Deployment.}''} Such agreements would require credible monitoring and verification systems to be effective. Developing strong verification mechanisms in the narrow context of export controls provides an opportunity to test the mechanisms that could support broader AI agreements in the future, while still providing immediate value by strengthening U.S. export enforcement.

Table~\ref{tab:summary} summarizes all the mechanisms discussed in this paper. See the \hyperref[appendix-methodology]{Appendix} for details on ratings.

\begin{table}[H]
\centering
\caption{Summary of the verification mechanisms discussed in this paper. See the \hyperref[appendix-methodology]{Appendix} for details on ratings. ``Novel'' denotes mechanisms not currently used for export control verification but implementable using existing frameworks and/or technologies.}
\label{tab:summary}
\scriptsize
\setlength{\tabcolsep}{4pt}
\begin{tabular}{@{}>{\raggedright\arraybackslash}p{2.7cm}>{\raggedright\arraybackslash}p{4.2cm}>{\raggedright\arraybackslash}p{2.3cm}>{\raggedright\arraybackslash}p{2.0cm}>{\raggedright\arraybackslash}p{2.2cm}@{}}
\toprule
Mechanism & Description & Maturity & Effectiveness & Access Required \\
\midrule
\multicolumn{5}{@{}l}{\textbf{\hyperref[end-location-verification]{End-Location Verification}}} \\
\addlinespace[2pt]
Export Documentation Checks & Tracing chips through export documentation (e.g.\ commercial invoice and bill of lading) & Relatively Established & Low & Not Invasive \\
\addlinespace[2pt]
On-Site Inspections & Physically visiting the chips on location and visually checking which chips are there & Established & High & Invasive \\
\addlinespace[2pt]
Remote Video Inspections & Conducting video inspection of data centers & Relatively Established & Medium & Relatively Invasive (only for setup) \\
\addlinespace[2pt]
Auditor-Directed Video Walkthroughs & Conducting interactive, auditor-directed remote walkthroughs of data-center facilities to verify the physical location of restricted chips & Relatively Established & Medium & Relatively Invasive \\
\addlinespace[2pt]
Random Return Requests & Issuing short-notice chip-return requests to end users & Novel & High & Relatively Invasive \\
\addlinespace[2pt]
Delay-Based Location Verification & Using response delay between chips and trusted landmark servers to verify location & Novel & High & Not Invasive \\
\addlinespace[3pt]
\midrule
\multicolumn{5}{@{}l}{\textbf{\hyperref[end-user-verification]{End-User Verification}}} \\
\addlinespace[2pt]
Enhanced Know-Your-Customer (KYC) Checks & Requiring companies to conduct robust KYC & Established & Medium & Not Invasive \\
\addlinespace[2pt]
KYC Compliance Audits & Confirming via independent review by accredited auditors that exporters have conducted KYC checks in line with BIS guidance & Relatively Established & Medium & Not Invasive \\
\addlinespace[3pt]
\midrule
\multicolumn{5}{@{}l}{\textbf{\hyperref[end-use-verification]{End-Use Verification}}} \\
\addlinespace[2pt]
End-Use and Line-of-Business Cross-Checks & Cross-checking declared use of compute with line of business & Relatively Established & Low & Not Invasive \\
\addlinespace[2pt]
Compute Provisioning Audits & Auditing computing capacity provisioned by cloud providers to customers & Novel & Medium & Relatively Invasive \\
\bottomrule
\end{tabular}
\end{table}

\section{End-Location Verification}\label{end-location-verification}

A core challenge in enforcing export controls on advanced AI chips is verifying that these items remain in their authorized destinations after export. Once chips leave the exporter's direct control, they can be resold, transferred, or integrated into larger clusters. Chip diversion not only undermines the effectiveness of export controls but also complicates the verification of compliance in potential future international AI governance efforts, since diverted compute cannot be easily tracked or reclaimed.

This section explores several approaches to verifying the location of AI chips, but no single tool is likely to be impossible to evade. The most effective option is to adopt a layered verification framework. For instance, delay-based location verification could serve as a continuous first line of detection, with remote inspections or auditor-directed walkthroughs providing periodic corroboration, and on-site inspections reserved for high-risk exports or cases where anomalies arise. Policymakers and auditors may combine these options in different ways depending on their priorities, risk tolerance, and available resources.

\subsection{Export Documentation Checks}\label{export-documentation-checks}

\emph{\textbf{Summary:} Trace a random sample of chips through the entire documentation chain (from manufacturer to exporter to consignee to final facility), including contracts, bills of lading, internal transfer forms, shipping manifests, transport logs, and receiving-facility intake records. This could be performed by accredited auditors, a company's internal compliance team, or BIS.}

By tracing the chain of custody for a random sample of exported chips, auditors can help verify their final location and assess a company's compliance with the EAR. This mechanism is similar to chain-of-custody tracing, an established method in the sustainable-commodities sector whereby certification schemes attest that every actor in the supply chain complies with certain standards.\footnote{SGS, ``\href{https://www.sgs.com/en/services/eu-deforestation-regulation-eudr-chain-of-custody-verification}{EU Deforestation Regulation (EUDR) Chain of Custody Verification.}''; SGS, ``\href{https://www.sgs.com/en/services/iscc-plus-certification}{ISCC PLUS Certification.}''}

Several actors may appear in the chain of custody. Some end customers purchase chips directly from AI chip designers like NVIDIA, some from server builders such as Super Micro, Dell, Lenovo, Gigabyte, or HPE, and some through distributors and resellers. Distributors rarely sell directly to end customers; instead, they typically partner with resellers. There are only a few major distributors---notably TD Synnex, Ingram Micro, and Arrow Electronics---while the reseller network is much more diffuse, composed of numerous smaller firms operating worldwide.\footnote{Grunewald, ``\href{https://www.the-substrate.net/p/how-banned-ai-chips-end-up-in-china}{How banned AI chips end up in China.}''}

\begin{figure}[htbp]
\centering
\begin{tikzpicture}[
  box/.style={draw=black!70, thick, rounded corners=2pt, fill=black!8,
              minimum height=2.6em, inner xsep=10pt, font=\small\bfseries},
  flow/.style={-{Stealth[length=2.4mm]}, densely dotted, thick, black!70},
  direct/.style={-{Stealth[length=2.4mm]}, thick, black},
  lab/.style={font=\small\itshape, black}
]
\node[box] (nvidia)  at (0,0)    {NVIDIA};
\node[box] (server)  at (5.6,0)  {Server builders};
\node[box] (endc)    at (11.2,0) {End customers};
\node[box] (dist)    at (2.4,-2.2) {Distributors};
\node[box] (resell)  at (7.6,-2.2) {Resellers};

\draw[flow] (nvidia) -- (server);
\draw[flow] (server) -- (endc);
\draw[direct] (nvidia.north) |- (5.6,1.3) node[lab, above, pos=0.75] {Direct sale} -| (endc.north);
\draw[flow] (nvidia.south) |- (dist.west);
\draw[flow] (dist) -- (resell);
\draw[flow] (resell.east) -| node[lab, right, pos=0.85] {Via reseller} (endc.south);
\end{tikzpicture}
\caption{NVIDIA's sales channels for AI chips.}
\label{fig:sales-channels}
\end{figure}
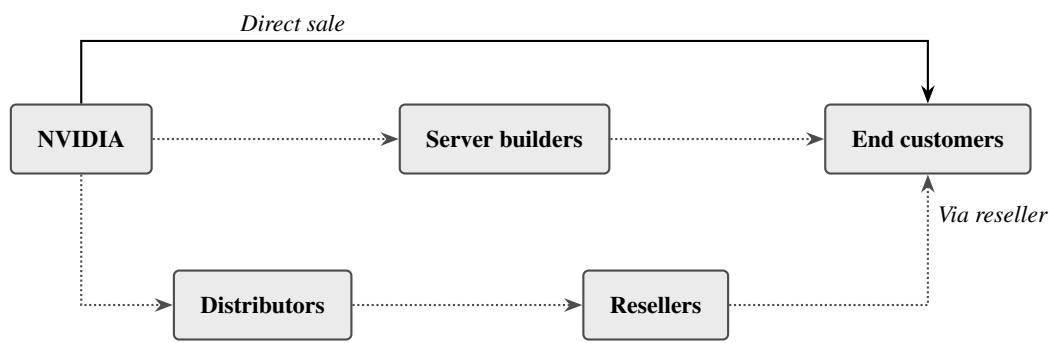

Because tracing every transaction in this network would involve multiple jurisdictions and numerous intermediaries, the scope of this mechanism must be narrowly defined, as following the full report trail of every chip would simply not be feasible, particularly given the difficulties of establishing liability once chips are resold by third parties.

\begin{tcolorbox}[breakable,colback=black!4,colframe=black!55,boxrule=0.6pt,arc=1.5pt,left=8pt,right=8pt,top=8pt,bottom=8pt,title={\textbf{Who is liable after chips are resold?}},fonttitle=\small,coltitle=white,colbacktitle=black!55]
\small
Under the current liability regime, NVIDIA and server builders are not responsible for what happens once chips are resold by distributors or resellers unless there is ``knowledge''\footnotemark{} that the reseller intends to divert the chips to unauthorized destinations or purposes. BIS's KYC guidance\footnotemark{} clarifies what this knowledge standard demands in practice: ``red flags'' in a transaction trigger a duty to inquire, and exporters cannot ``self-blind'' by avoiding having access to relevant information.

\medskip
Foreign firms and individuals reselling AI chips are also liable under the EAR: ``the law follows the goods''.\footnotemark{} BIS enforces liability against foreign actors through civil penalties, which it can impose unilaterally, and through criminal arrests, pursued separately via indictment and extradition by the Department of Justice. Tracking the chain of custody of chips once they become the property of a foreign company may be impractical without formal cooperation frameworks between the U.S. and foreign governments, as it would require direct access to company records and potentially on-site inspections abroad.
\end{tcolorbox}
\addtocounter{footnote}{-2}
\footnotetext{Code of Federal Regulations, ``\href{https://www.ecfr.gov/current/title-15/subtitle-B/chapter-VII/subchapter-C/part-764/section-764.2}{\S~764.2 Violations.}''}
\stepcounter{footnote}
\footnotetext{U.S. Department of Commerce, ``\href{https://www.bis.gov/node/1533}{Supplement No. 3 to Part 732---BIS's ``Know Your Customer'' Guidance and Red Flags.}''}
\stepcounter{footnote}
\footnotetext{U.S. Department of Commerce et al., ``\href{https://www.justice.gov/archives/opa/media/1341411/dl?inline}{Obligations of foreign-based persons to comply with U.S. sanctions and export control laws.}''}

Given these limitations, the most practical purpose of export documentation checks would be to confirm that U.S. companies and their subsidiaries overseas are not directly selling chips to prohibited customers or for restricted end uses.\footnote{Considering the aforementioned indirect liability challenges, this method would not cover indirect sales.} It is also important to note that this mechanism has significant vulnerabilities: documents can be forged, and companies can conceal information through obscure corporate structures. Combining documentation checks with enhanced KYC checks (see ``\hyperref[enhanced-kyc-checks]{Enhanced Know-Your-Customer (KYC) Checks}'') can help address some of these weaknesses, but as a stand-alone mechanism it would likely serve as a meaningful verification tool only against actors with low to moderate sophistication.

Implementation steps:

\begin{enumerate}
\def\labelenumi{\arabic{enumi}.}
\item
  \textbf{Export Licensing and Shipment Validation:} To confirm that the export was properly authorized, the exporter, an accredited auditor, or BIS reviews BIS license or license-exception documentation; Export Control Classification Number (ECCN), Electronic Export Information (EEI), and Automated Export System Internal Transaction Number (AES ITN); commercial invoice; purchase order; and end-use/end-user statement, cross-checking all parties, quantities, and destinations against the authorization. Much of the document review process proposed in this paper can be automated using LLMs; BIS could develop an online platform where relevant documents are uploaded and reviewed automatically---we refer to this proposed platform as the BIS compliance platform.
\end{enumerate}

\emph{Performed by: accredited auditor, the exporter's internal due-diligence team, or BIS. BIS is authorized}\footnote{Code of Federal Regulations, ``\href{https://www.ecfr.gov/current/title-15/subtitle-B/chapter-VII/subchapter-C/part-758/section-758.7}{§ 758.7 Authorities of the Bureau of Industry and Security, Office of Export Enforcement (OEE).}''} \emph{to conduct this type of review, although the extent to which it is systematically conducted is unclear.}

\begin{enumerate}
\def\labelenumi{\arabic{enumi}.}
\setcounter{enumi}{1}
\item
  \textbf{Customs and Logistics Reconciliation:} The exporter, an accredited auditor or BIS verifies that the goods moved as declared by matching the license application with the following documents:

  \begin{itemize}
  \item
    Carrier tracking data: shipment-movement information generated by a carrier's logistics system during transport, such as container or package tracking numbers.
  \item
    Air waybill or bill of lading\footnote{International Trade Administration, ``\href{https://www.trade.gov/common-export-documents}{Common Export Documents.}''}: documents accompanying goods shipped by international air freight or by ocean or other transport, respectively, which provide detailed information about the contents of the shipment and the parties involved.
  \item
    Warehouse receiving reports and inventory logs: records that confirm shipments arrived at licensed consignees and that quantities match manifests.
  \end{itemize}
\end{enumerate}

\emph{Performed by: accredited auditor, the exporter's internal due-diligence team, or BIS.}\footnote{Under the EAR, the exporter of record (or its freight forwarder) must electronically file shipment details---known as Electronic Export Information (EEI)---through the Automated Export System (AES) prior to export. Upon acceptance, the system issues an Internal Transaction Number (ITN) as proof of filing. These AES filings are accessible to BIS, the Census Bureau, and Customs and Border Protection (CBP) for enforcement and compliance purposes but are not publicly disclosed for commercial and national-security reasons. If accredited auditors are involved, BIS could establish an information-sharing platform enabling them to access relevant AES data securely.} \emph{BIS is authorized}\footnote{Code of Federal Regulations, ``\href{https://www.ecfr.gov/current/title-15/subtitle-B/chapter-VII/subchapter-C/part-758/section-758.7}{§ 758.7 Authorities of the Bureau of Industry and Security, Office of Export Enforcement (OEE).}''} \emph{to conduct this type of review, although the extent to which it is systematically conducted is unclear.}

\subsection{On-Site Inspections}\label{on-site-inspections}

\emph{\textbf{Summary:} Conduct on-site inspections of data-center facilities to verify the physical location of restricted chips by visually confirming serial numbers, reviewing inventory records, and cross-checking them against numbers reported to BIS. Inspections could be scheduled or unannounced and performed by BIS or accredited auditors based on risk level and compliance history.}

Implementation steps:

\begin{enumerate}
\def\labelenumi{\arabic{enumi}.}
\item
  \textbf{Pre-Inspection Document Review:} Prior to the inspection, the company uploads all relevant documentation either to the BIS online compliance platform or directly to an accredited auditor. This includes the export license, end-use statement, inventory list with serial numbers, facility layout or rack map, and shipping and receiving logs confirming the chips' arrival. The company should also provide in advance the name and credentials of all representatives who will accompany or guide inspectors during the visit. The auditor reviews these materials to understand what must be verified during the session. It is worth noting that this step would be most effective alongside a reliable export documentation regime, as inspectors can only verify that the chips that are supposed to be at a location are indeed there if they have a record of which chips are supposed to be there.\footnote{That said, even if such a regime doesn't exist, inspections can still be useful. For example, the inspector may have a vague sense that a very large number of chips shouldn't be at the location, even if the inspector doesn't know the exact number or serial numbers of those chips. If the inspector then visits a facility that cannot, or does not, house that many chips, it is likely that a violation has occurred. This would warrant further investigation.}
\end{enumerate}

\emph{Performed by}: \emph{BIS or accredited auditor.}

\begin{enumerate}
\def\labelenumi{\arabic{enumi}.}
\setcounter{enumi}{1}
\item
  \textbf{On-Site Inventory Verification}: Inspectors confirm that the number of chips and serial numbers at the facility match those reported to BIS. For this cross-check, BIS could rely on records it already holds---quantities and end users listed on export license applications and in export filings---though these capture cumulative shipments rather than current on-site inventories, and existing recordkeeping requirements under 15 C.F.R. § 762.2(a) do not include inventory logs. Alternatively, BIS could expand these obligations to require inventory records as it already does for Data Center Validated End Users, which must maintain inventory logs and submit semi-annual reports.\footnote{Industry and Security Bureau, ``\href{https://www.federalregister.gov/documents/2024/10/02/2024-22587/expansion-of-validated-end-user-authorization-data-center-validated-end-user-authorization}{Expansion of Validated End User Authorization: Data Center Validated End User Authorization.}''} This step would involve physically checking the device serial numbers of a random sample of chips and reviewing rack-level asset inventories or equipment management databases.\footnote{It may be possible to forge serial numbers on chips and racks. A stronger approach, though a more time- and resource-intensive one, would be to use serial-number checks as a screening step, with cryptographic attestation of a powered-on chip as confirmation.}
\end{enumerate}

The 2026 indictment of three men affiliated with Super Micro for allegedly conspiring to smuggle chips to China---including one executive and another senior employee---illustrates how on-site inspections can be forged, and what inspectors must guard against. The scheme involved staging non-working physical server replicas repackaged with manufacturer labels and serial-number stickers to substitute for equipment that had already been illegally shipped to China. One of the defendants impersonated an assistant from the facility's local law firm during the inspection.\footnote{United States District Court Southern District Of New York, ``\href{https://storage.courtlistener.com/recap/gov.uscourts.nysd.660030/gov.uscourts.nysd.660030.2.0.pdf}{UNITED STATES OF AMERICA V. YIH-SHYAN "Wally" LIAW, RUEI-TSANG "Steven" CHANG, and TING-WEI "Willy" SUN.}''} Upon arrival, inspectors should verify the credentials of the company representatives present against those submitted in advance. They should also physically probe equipment to verify it is operational, and examine labels and serial-number stickers for signs of tampering.\\
\emph{Performed by: BIS or accredited auditor.}

\begin{enumerate}
\def\labelenumi{\arabic{enumi}.}
\setcounter{enumi}{2}
\item
  \textbf{Photographic and Digital Evidence Collection:} During the inspection, auditors capture timestamped, geotagged photographs of the verified equipment, along with screenshots of the corresponding inventory records. These materials are uploaded to an online database managed by BIS to store the results of inspections and linked to the facility's compliance record.
\end{enumerate}

\emph{Performed by: BIS or accredited auditor.}

\begin{enumerate}
\def\labelenumi{\arabic{enumi}.}
\setcounter{enumi}{3}
\item
  \textbf{Final Verification Report:} The auditor or BIS prepares an \emph{On-Site Inspection Verification Report} summarizing: the number and serial numbers of chips verified, any discrepancies found, explanations or proposed corrective actions, and recommendations for further review or enforcement. The report is uploaded to the BIS online compliance platform and may serve as input for follow-up inspections or license renewals.
\end{enumerate}

\emph{Performed by: BIS or accredited auditor.}

\subsection{Remote Video Inspections}\label{remote-video-inspections}

\emph{\textbf{Summary:} Conduct remote video verifications of data-center deployments to confirm the location of restricted chips. This could be performed by BIS or accredited auditors.}

Remote video inspections offer a lower-cost alternative to on-site visits, allowing BIS or accredited auditors to verify the physical location of chips without needing direct access to the facility. Unlike on-site inspections, auditors conduct these checks through secure live-video sessions, during which company staff can show the facility exterior, server rooms, rack labels, and device serial numbers, and provide screen captures from system dashboards confirming the chips' presence and configuration.

Similar inspections have been used for other regimes, for example, the International Atomic Energy Agency's Next-Generation Surveillance System, which deploys cameras housed in tamper-indicating containers to verify nuclear-material locations without continuous on-site presence, and the UN Special Commission program in Iraq, which installed remote-controlled cameras at missile-engine test stands to confirm that no prohibited activities were taking place.\footnote{Fournier, ``\href{https://www.iaea.org/newscenter/news/surveying-safeguarded-material-24/7}{Surveying Safeguarded Material 24/7.}''; United Nations, ``\href{https://www.un.org/Depts/unscom/sres26910.htm}{Sixth report of the Executive Chairman of the Special Commission.}''} Both systems used fixed, permanently installed cameras rather than conducting walkthroughs or live interactive sessions, functioning more as continuous monitoring tools.\footnote{Fournier, ``\href{https://www.iaea.org/newscenter/news/surveying-safeguarded-material-24/7}{Surveying Safeguarded Material 24/7.}''} Because advanced AI chips are essential for training and deploying frontier AI models, they are strategic assets for national security and may warrant a similar level of monitoring.

While remote verifications are less robust than in-person inspections, they would still provide a practical and lower-cost layer of verification if designed with anti-spoofing safeguards, potentially serving as a follow-up to previous on-site inspections. AI-enabled spoofing is a significant threat in this context, as real-time deepfake-video generation could replace a live camera feed with a synthetic one that appears to show a compliant facility.

To minimize the risk of spoofing, inspections should be conducted over auditor-controlled, end-to-end encrypted streams and combined with randomized auditor requests during the session, such as asking staff to perform unscripted actions. An additional safeguard could be to implement a system similar to Laser Curtain for Containment and Tracking (LCCT), which is used in nuclear verification.\footnote{Sequeira et al., ``\href{https://resources.inmm.org/sites/default/files/2021-09/a498.pdf}{Laser curtain for containment and tracking.}''} In this system, video cameras are combined with laser scanners to create a virtual perimeter around certain guarded items, continuously monitoring the area for any unauthorized movement. Because this approach uses active physical sensing rather than video feeds alone, anyone trying to spoof a remote inspection would have to manipulate several independent monitoring systems at once, which would be harder to do and more likely to be detected.

Still, it is important to keep in mind that on-site and remote inspections should always be part of a wider verification system. IAEA's early safeguards approach, which relied heavily on mechanistic, facility-based inspections, contributed to its failure to detect Iraq's clandestine nuclear-weapons program in the late 1980s and early 1990s.\footnote{Baute, ``\href{https://www.iaea.org/sites/default/files/publications/magazines/bulletin/bull46-1/46102486468.pdf}{Timeline IRAQ: Challenges \& Lessons Learned from Nuclear Inspections.}''} Following this failure, the institution moved toward the state-level approach, which calibrates verification based on a broad range of information about a state's nuclear capabilities and tailors safeguards accordingly. Similarly, remote inspections for export control enforcement should be embedded in a multi-pronged verification system designed to detect evasion with a high enough likelihood that, at minimum, it imposes substantial costs or friction on bad actors, rather than simply confirming declared information.

Implementation steps:

\begin{enumerate}
\def\labelenumi{\arabic{enumi}.}
\item
  \textbf{Pre-Inspection Document Review:} See \hyperref[on-site-inspections]{On-Site Inspections}.
\item
  \textbf{Live Remote Verification:} On the scheduled day, the company connects via a secure video platform (for example, a BIS-approved remote-inspection portal). The video feed should come from a camera system that embeds geolocation and timestamp metadata and includes a camera identity certificate.\footnote{Ideally, data centers would use IAEA-grade (International Atomic Energy Agency, ``\href{https://www.iaea.org/newscenter/multimedia/photoessays/equipment-used-safeguards}{Equipment Used in Safeguards.}'') cameras---purpose-built for robust, tamper-proof verification---but these are not commercially available. Among off-the-shelf alternatives, the Axis Q6078-E (Axis Communications, ``\href{https://www.axis.com/products/axis-q6078-e}{Product support for AXIS Q6078-E PTZ Camera.}'') camera appears suitable for this purpose. Available documentation (Axis Communications, ``\href{https://www.axis.com/dam/public/22/dc/ba/datasheet-axis-q6078-e-ptz-camera-en-US-487394.pdf}{AXIS Q6078-E PTZ Camera.}'') confirms that the model includes trusted platform modules (TPMs) and digital certificates; however, we could find no information confirming whether it also embeds geolocation or timestamp metadata.} The identity certificate is a cryptographic credential stored in the camera's hardware, allowing the auditor to verify that the footage originates exactly from that specific device, thereby ensuring the device has not been spoofed or the footage pre-recorded. Additionally, each camera should be positioned within the field of view of at least one other camera, such that any attempt to tamper with or obstruct a device would itself be recorded by its counterpart.
\end{enumerate}

\emph{Performed by: BIS or accredited auditor.}

3. \textbf{Photographic and Digital Evidence Collection:} See \hyperref[on-site-inspections]{On-Site Inspections}.

4. \textbf{Final Verification Report:} See \hyperref[on-site-inspections]{On-Site Inspections}.

\subsection{Auditor-Directed Video Walkthroughs}\label{auditor-directed-video-walkthroughs}

\emph{\textbf{Summary:} Conduct interactive, auditor-directed remote walkthroughs of data-center facilities to verify the physical location of restricted chips. This can be performed by BIS or accredited auditors.}

Similar to the remote video inspections detailed above, auditor-directed video walkthroughs verify the location of chips via a secure live-video feed. But instead of providing continuous monitoring by relying on fixed cameras, these walkthroughs are conducted by staff members of the inspected facility who move through the data center with a camera, following instructions from the auditor. To avoid AI spoofing, the auditor can issue unpredictable instructions to the staff during the inspection. This mechanism can be used to verify the location of chips in facilities where installing fixed cameras is not feasible, or as a follow-up to red flags detected through other verification methods.

\emph{Implementation steps}:

\begin{enumerate}
\def\labelenumi{\arabic{enumi}.}
\item
  \textbf{Pre-Inspection Document Review:} See \hyperref[on-site-inspections]{On-Site Inspections}.
\item
  \textbf{Auditor-Directed Navigation:} On the scheduled day, the company connects via a secure video platform. During the session, company staff show the facility exterior, server rooms, rack labels, and device serial numbers, and provide screen captures from system dashboards confirming the chips' presence and configuration. The auditor directs the walkthrough in real time, instructing on-site staff to navigate to specific locations within the facility. The auditor may also ask staff to open particular racks or servers to verify authenticity or to display a random sample of chips by serial number.
\end{enumerate}

\emph{Performed by: BIS or accredited auditor.}

\begin{enumerate}
\def\labelenumi{\arabic{enumi}.}
\setcounter{enumi}{2}
\item
  \textbf{Live-Feed Verification:} During the session, the auditor asks the staff of the inspected facility to perform certain tasks to confirm that the video is being transmitted live, such as holding up a code given by the auditor or showing a specific rack. This helps prevent the use of manipulated or AI-generated video.
\end{enumerate}

\emph{Performed by: BIS or accredited auditor.}

\begin{enumerate}
\def\labelenumi{\arabic{enumi}.}
\setcounter{enumi}{3}
\item
  \textbf{Cryptographic Timestamping and Metadata:} The video feed used during the walkthrough should embed cryptographically signed timestamps and geolocation metadata. This does not completely prevent spoofing, but it raises the bar for manipulation by requiring the adversary to forge metadata in real time without disrupting the session. \emph{Performed by: company, with equipment meeting BIS-approved specifications.}
\item
  \textbf{Cross-Referencing with Other Verification Data}: Information observed during the walkthrough---such as the number of racks, serial numbers of inspected chips, and facility layout---should be cross-referenced with available records (e.g. export documentation and inventory records). Discrepancies between the walkthrough observations and other data sources trigger further investigation.
\end{enumerate}

\emph{Performed by: BIS or accredited auditor.}

\subsection{Random Return Requests}\label{random-return-requests}

\emph{\textbf{Summary:} Issue random return requests for specific chips (by serial number) at short notice to confirm with high probability that they are not within prohibited jurisdictions. Failure to return a requested chip within the set timeframe would trigger further investigation or an on-site inspection.}

Another alternative to on-site inspections is to require random return requests for specific chips (by serial number) at short notice to confirm that they remain at the authorized end location.\footnote{Fist and Grunewald, ``\href{https://www.cnas.org/publications/reports/preventing-ai-chip-smuggling-to-china}{Preventing AI Chip Smuggling to China.}''} This reduces the cost for the inspector by removing the need to travel to the chips' location. It may also be cheaper for the chip owner, who does not need to arrange for inspectors to visit the facilities and accompany them when they are on site.

Removing graphics processing units (GPUs) from active servers on demand is not likely to be a significant burden to chip owners. Data centers already maintain processes for swapping out chips that fail or require maintenance, and an operational study of NVIDIA A100 and H100 GPUs at the National Center for Supercomputing Applications observed a physical replacement rate of approximately 0.45\% per year across a fleet of 1,056 export controlled GPUs---meaning that for every 2,000 such chips, roughly nine are already swapped out each year due to hardware and memory failures alone.\footnote{Fist and Grunewald, ``\href{https://www.cnas.org/publications/reports/preventing-ai-chip-smuggling-to-china}{Preventing AI Chip Smuggling to China.}''; Cui et al., ``\href{https://arxiv.org/abs/2503.11901}{Story of Two GPUs: Characterizing the Resilience of Hopper H100 and Ampere A100 GPUs.}''} A random-return program would add only a small fraction to this existing swap activity.\footnote{Fist and Grunewald, ``\href{https://www.cnas.org/publications/reports/preventing-ai-chip-smuggling-to-china}{Preventing AI Chip Smuggling to China.}''} To further minimize disruption, BIS could maintain a reserve stock of controlled chips that can be temporarily swapped in while the selected chip is in transit for inspection.\footnote{Fist and Grunewald, ``\href{https://www.cnas.org/publications/reports/preventing-ai-chip-smuggling-to-china}{Preventing AI Chip Smuggling to China.}''}

To make this mechanism operational, BIS could establish a centralized registry of chip ownership from which it can select random IDs for return requests. Ideally, the registry would cover both U.S. exports and re-exports from third countries, but achieving that scope would require a degree of cooperation from re-exporters and foreign governments that may not be feasible across all jurisdictions. Beyond enabling this verification mechanism, such a registry would be beneficial because it would create a single source of truth for cross-checks that could help identify inconsistencies across reports and identify patterns of chip diversion. Chips could be required to be shipped within a short timeframe (e.g. 24--48 hours) to a nearby U.S. embassy, consulate, or U.S. Commercial Service office. The tight deadline helps confirm that the chips were already in-country at the time of the request. Failures to comply, late arrivals, or evidence of tampering would trigger further investigation or an on-site audit.

Receiving, inventorying, and verifying AI chips would require the relevant U.S. diplomatic and commercial posts to have dedicated secure storage space, standardized intake procedures, and personnel trained both in chip identification and serial-number verification. Staff should also be equipped to recognize spoofing attempts---for instance, serial-number swapping, in which a chip's identifier is stripped and affixed to a different chip to make it appear compliant, or the submission of replicas bearing the correct identifiers in place of the genuine chip. This would represent a meaningful administrative investment. One way to reduce the cost would be to limit initial deployment to a small number of locations at a high risk of diverting chips, such as Singapore, Thailand, and Malaysia.

Implementation steps:

\begin{enumerate}
\def\labelenumi{\arabic{enumi}.}
\item
  \textbf{Chip-Registry Establishment:} BIS creates a centralized registry of chip serial numbers and ownership records.
\item
  \textbf{Random Sampling:} BIS or accredited auditors generate a random list of chip serial numbers for return verification.
\item
  \textbf{Return Request:} The selected end users receive a formal request to ship the specified chips back to an authorized verification facility (e.g. U.S. embassy, consulate, or regional U.S. Commercial Service office). BIS could require end users to ship the chips within a short timeframe (e.g. 24--48 hours) and to use an established shipping company like FedEx, DHL or UPS to reduce the risk that the company could ship the chip from a non-authorized location.
\item
  \textbf{Return Confirmation:} Upon receipt, the chips' serial numbers and identifiers are verified against export and inventory records, and once confirmed, BIS returns the chips to the end user.
\end{enumerate}

\subsection{Delay-Based Location Verification}\label{delay-based-location-verification}

\emph{\textbf{Summary:} Use the built-in features of advanced AI chips to verify their physical location in near-real-time by measuring ping response times to trusted landmark servers.}

Ping-based location verification would use the built-in attestation\footnote{Attestation is the process by which modern AI chips verify their identity using a unique cryptographic key embedded in each device (Brass and Aarne, ``\href{https://www.iaps.ai/research/location-verification-for-ai-chips}{Location Verification for AI Chips.}'').} features of advanced AI chips to confirm their physical location.\footnote{Brass and Aarne, ``\href{https://www.iaps.ai/research/location-verification-for-ai-chips}{Location Verification for AI Chips.}''} This method provides near--real-time visibility of where chips are operating by measuring the response time between a chip and a trusted landmark server at a known location.\footnote{Brass and Aarne, ``\href{https://www.iaps.ai/research/location-verification-for-ai-chips}{Location Verification for AI Chips.}''} The overall cost of implementation at scale would be relatively low compared to the profit margins of major chip manufacturers---under \$1 million for firmware development and between \$2.5 million and \$12.5 million annually for a landmark server network.\footnote{Brass and Aarne, ``\href{https://www.iaps.ai/research/location-verification-for-ai-chips}{Location Verification for AI Chips.}''} It is worth noting, however, that this mechanism only becomes operational once chips are installed and networked in a data center. During shipping and warehousing, the chips cannot perform cryptographic challenge-response communication with landmark servers. This means the method provides limited visibility during the period between export and deployment---any diversion during transit would not be detectable.

Advanced AI chips contain unique cryptographic identifiers, allowing them to authenticate themselves and respond to pings (empty messages used to measure round-trip time) from landmark servers. By measuring the latency of these responses, it is possible to determine whether the chip is within a feasible distance of the landmark. A rudimentary version of this approach has already been successfully prototyped for NVIDIA H100 chips, and NVIDIA has since confirmed that it is actively developing location-verification technology based on this principle, using communication delays with NVIDIA-run servers to estimate chip location. This mechanism will reportedly debut on Blackwell chips.\footnote{NVIDIA Newsroom, ``\href{https://blogs.nvidia.com/blog/optional-data-center-fleet-management-software/}{Opt-In NVIDIA Software Enables Data Center Fleet Management.}''; Nellis and Martina, ``\href{https://www.reuters.com/business/nvidia-builds-location-verification-tech-that-could-help-fight-chip-smuggling-2025-12-10/}{Nvidia builds location verification tech that could help fight chip smuggling.}''}

However, this location-verification method can be evaded: malicious actors could artificially inflate ping response times---for instance by routing traffic along a circuitous path---to obscure a chip's location, or reduce them by exploiting dark fiber and high-speed interconnects to make a chip appear to be operating in a different location entirely.\footnote{Brass and Aarne, ``\href{https://www.iaps.ai/research/location-verification-for-ai-chips}{Location Verification for AI Chips.}''; Nellis and Martina, ``\href{https://www.reuters.com/business/nvidia-builds-location-verification-tech-that-could-help-fight-chip-smuggling-2025-12-10/}{Nvidia builds location verification tech that could help fight chip smuggling.}''; NVIDIA Newsroom, ``\href{https://blogs.nvidia.com/blog/optional-data-center-fleet-management-software/}{Opt-In NVIDIA Software Enables Data Center Fleet Management.}''} Landmark servers themselves are also a potential point of compromise: an adversary that gains control over one or more landmarks can directly spoof communication-travel-time measurements without relying on ``man-in-the-middle network manipulation''.\footnote{Brass and Aarne, ``\href{https://www.iaps.ai/research/location-verification-for-ai-chips}{Location Verification for AI Chips.}''; Nellis and Martina, ``\href{https://www.reuters.com/business/nvidia-builds-location-verification-tech-that-could-help-fight-chip-smuggling-2025-12-10/}{Nvidia builds location verification tech that could help fight chip smuggling.}''; NVIDIA Newsroom, ``\href{https://blogs.nvidia.com/blog/optional-data-center-fleet-management-software/}{Opt-In NVIDIA Software Enables Data Center Fleet Management.}''} Measures to mitigate these evasion attempts include maintaining a centralized chip registry that requires chips to be registered at specific locations and deploying dense networks of landmark servers near jurisdictions considered at high risk of diverting chips.\footnote{Brass and Aarne, ``\href{https://www.iaps.ai/research/location-verification-for-ai-chips}{Location Verification for AI Chips.}''; Nellis and Martina, ``\href{https://www.reuters.com/business/nvidia-builds-location-verification-tech-that-could-help-fight-chip-smuggling-2025-12-10/}{Nvidia builds location verification tech that could help fight chip smuggling.}''; NVIDIA Newsroom, ``\href{https://blogs.nvidia.com/blog/optional-data-center-fleet-management-software/}{Opt-In NVIDIA Software Enables Data Center Fleet Management.}''}

Location-verification technology is also being considered in legislation as a mechanism to counter chip smuggling. The Chip Security Act, a bipartisan bill that would compel the Secretary of Commerce to require location-verification mechanisms for advanced AI chips that are exported, passed the House Foreign Affairs Committee unanimously in March 2026.\footnote{Huizenga, ``\href{https://www.congress.gov/bill/119th-congress/house-bill/3447/text}{H.R.3447 - Chip Security Act.}''; Select Committee on China, ``\href{https://chinaselectcommittee.house.gov/media/press-releases/house-committee-passes-chip-security-act}{House Committee Passes Chip Security Act.}''} The bill requires chips to be outfitted with ``mechanisms that implement location verification, using techniques that are feasible and appropriate on such date of enactment'', meaning delay-based location verification is one approach that could be used.\footnote{Huizenga, ``\href{https://www.congress.gov/bill/119th-congress/house-bill/3447/text}{H.R.3447 - Chip Security Act.}''}

Implementation steps:

\begin{enumerate}
\def\labelenumi{\arabic{enumi}.}
\item
  \textbf{Firmware and Software Update:} Chip manufacturers could deploy a firmware update that enables secure, automated delay-based location verification. This requires providing unique secret keys to each chip, storing them securely via a trusted platform module (TPM) or similar solution, and implementing a limited interface with the rest of the chip for cryptographic challenge-response communication with landmark servers. It would also require adding or updating on-chip software to handle communication with landmark servers.\\
  \emph{Performed by: chip manufacturers (e.g. NVIDIA or AMD).}
\item
  \textbf{Establishment of Landmark Server Networks:} A network of trusted landmark servers would be established in or near major data-center hubs. Each landmark can verify chip pings and record results.\\
  \emph{Performed by: chip manufacturers (e.g.} NVIDIA or AMD\emph{).}
\item
  \textbf{Periodic Verification Requests:} BIS can periodically request exporters or end-users of controlled chips to initiate verification pings to the nearest landmark server.
\end{enumerate}

\emph{Performed by: exporters or end users.}

\section{End-User Verification}\label{end-user-verification}

Compliance with export controls largely depends on exporters knowing who the end users are. Verifying the identity of purchasers can be challenging, as entities seeking to evade controls often conceal ownership through shell companies or other intermediaries. Nonetheless, performing customer due diligence to ensure that buyers are not owned by or linked to entities of concern can help prevent violations of the EAR. In its KYC Guidance and Red Flags supplement,\footnote{Code of Federal Regulations, ``\href{https://www.ecfr.gov/current/title-15/subtitle-B/chapter-VII/subchapter-C/part-732/appendix-Supplement\%20No.\%203\%20to\%20Part\%20732}{Supplement No. 3 to Part 732---BIS\textquotesingle s ``Know Your Customer'' Guidance and Red Flags.}''} BIS outlines actions exporters can take to understand end users and end uses. Firms must watch for red flags indicating that an export may be intended for an unauthorized end user, such as orders that are inconsistent with the buyer's stated needs or requests for configurations unsuitable for the declared location. They must also avoid self-blinding, such as by instructing their sales staff to discourage customers from discussing the intended end use.\footnote{U.S. Department of Commerce, ``\href{https://www.bis.gov/node/1533}{Supplement No. 3 to Part 732---BIS\textquotesingle s "Know Your Customer" Guidance and Red Flags.}''} In addition, BIS introduced new reporting requirements for front-end chip fabricators that enhance due diligence by requiring quarterly reports to BIS detailing KYC information on certain integrated circuit designer customers.\footnote{Code of Federal Regulations, ``\href{https://www.ecfr.gov/current/title-15/subtitle-B/chapter-VII/subchapter-C/part-743/section-743.9}{§ 743.9 Reporting requirements for ``front-end fabricators'' producing ``applicable advanced logic integrated circuits'' for authorized integrated circuit designers.}''}

This section presents two mechanisms that policymakers could consider to strengthen end-user verification: Enhanced KYC checks, which would require exporters to conduct more rigorous due diligence on customers in defined high-risk cases, and KYC compliance audit, which would provide independent verification that those checks are being conducted properly.

\subsection{Enhanced KYC Checks}\label{enhanced-kyc-checks}

\emph{\textbf{Summary:} Require companies that export chips to conduct enhanced KYC checks for their customers in line with ``BIS's `Know Your Customer' Guidance and Red Flags}\footnote{Code of Federal Regulations, ``\href{https://www.ecfr.gov/current/title-15/subtitle-B/chapter-VII/subchapter-C/part-732/appendix-Supplement\%20No.\%203\%20to\%20Part\%20732}{Supplement No. 3 to Part 732---BIS\textquotesingle s ``Know Your Customer'' Guidance and Red Flags.}''} \emph{supplement. These reviews can be conducted by the company's internal compliance team or accredited auditors.}

Exporters already conduct KYC checks that are mostly limited to document and information collection to confirm that a company is not a shell or front company associated with restricted end users. These checks typically involve gathering general company information---business name, address, incorporation date, declared affiliations, and financial statements. In some cases, exporters may take additional due diligence measures, such as cross-checking that information against local business registries.\footnote{Grunewald, ``\href{https://www.the-substrate.net/p/how-banned-ai-chips-end-up-in-china}{How banned AI chips end up in China.}''} We propose that, for certain high-risk cases defined by BIS, these checks should be more thorough and include additional measures, as described in the steps below. High-risk cases can be defined by reference to BIS's KYC Guidance and Red Flags\footnote{Code of Federal Regulations, ``\href{https://www.ecfr.gov/current/title-15/subtitle-B/chapter-VII/subchapter-C/part-732/appendix-Supplement\%20No.\%203\%20to\%20Part\%20732}{Supplement No. 3 to Part 732---BIS\textquotesingle s ``Know Your Customer'' Guidance and Red Flags.}''} supplement: transactions where exporters identify abnormal circumstances suggesting an export may be destined for inappropriate end users or end uses (e.g. orders inconsistent with the purchaser's needs or business activities).\footnote{Code of Federal Regulations, ``\href{https://www.ecfr.gov/current/title-15/subtitle-B/chapter-VII/subchapter-C/part-732/appendix-Supplement\%20No.\%203\%20to\%20Part\%20732}{Supplement No. 3 to Part 732---BIS\textquotesingle s ``Know Your Customer'' Guidance and Red Flags.}''}

Implementation steps:

\begin{enumerate}
\def\labelenumi{\arabic{enumi}.}
\item
  \textbf{Document Collection:} Collect the following information or documents pertaining to the end user:

  \begin{itemize}
  \item
    Full legal name, registration number, and business address
  \item
    Corporate ownership and subsidiary structure (parent entities, beneficial owners, and shareholding percentages)
  \item
    Names of top executives, directors, and significant shareholders
  \item
    Intended end use (e.g. AI model training, cloud services, data analytics)
  \item
    Location of compute resources or facility addresses;
  \item
    Certifications and no-military or weapons of mass destruction (WMD)-use attestations.
  \end{itemize}
\end{enumerate}

\emph{Performed by: company's internal compliance team or accredited auditors.}

\begin{enumerate}
\def\labelenumi{\arabic{enumi}.}
\setcounter{enumi}{1}
\item
  \textbf{Screening and Risk Analysis:} The collected information should be screened to confirm that customers are not listed on the Entity List,\footnote{U.S. Department of Commerce, ``\href{https://www.bis.gov/regulations/ear/744\#section-744.16}{Part 744 - Control Policy: End-user and End-use Based.}''} the Office of Foreign Assets Control sanctions lists,\footnote{U.S. Department of Treasury, ``\href{https://ofac.treasury.gov/sanctions-list-search-tool}{Sanctions List Search Tool.}''} Denied Persons List,\footnote{U.S. Department of Commerce, ``\href{https://www.bis.gov/licensing/end-user-guidance/denied-persons-list-dpl}{Denied Persons List (DPL).}''} Unverified List,\footnote{U.S. Department of Commerce, ``\href{https://www.bis.gov/regulations/ear/744\#section-744.16}{Part 744 - Control Policy: End-user and End-use Based.}''} and the Military End-User List.\footnote{Code of Federal Regulations, ``\href{https://www.ecfr.gov/current/title-15/subtitle-B/chapter-VII/subchapter-C/part-744\#Supplement-No.-7-to-Part-744}{Part 744---control policy: end-user and end-use based.}''} Because this process is resource-intensive, a practical option is to use automated compliance platforms capable of performing supply chain risk analysis (e.g. SolidIntel,\footnote{SolidIntel, ``\href{https://solidintel.com}{SolidIntel.}''} AEB,\footnote{AEB, ``\href{https://www.aeb.com/en/compliance-screening/index.php}{Straightforward and secure restricted party screening.}''} and WireScreen\footnote{Wirescreen, ``\href{https://wirescreen.ai/}{WireScreen.}''}). The primary vulnerability of this approach is that the effectiveness of automated platforms depends on the underlying dataset---a platform with incomplete or outdated data may fail to identify sophisticated actors concealing their identity through complex corporate structures. BIS may eventually need to establish an accreditation system for platforms authorized to perform this service, ensuring they meet minimum standards for data coverage and analytical capability.
\end{enumerate}

\emph{Performed by: company's internal compliance team, accredited auditors, or accredited automated supply chain risk analysis platforms.}

\begin{enumerate}
\def\labelenumi{\arabic{enumi}.}
\setcounter{enumi}{2}
\item
  \textbf{KYC Attestation Form Submission via BIS Online Compliance Platform:} Once the review is complete, the company fills out a standardized KYC attestation form confirming that: the customer was screened against all relevant lists; beneficial ownership was verified; no red flags were identified, or, if present, they were resolved and documented. The form and supporting materials would be uploaded to a BIS online compliance platform.
\end{enumerate}

\subsection{KYC Compliance Audits}\label{kyc-compliance-audits}

\emph{\textbf{Summary:} Verify that the company performed KYC in accordance with BIS's KYC Guidance and Red Flags supplement. This can be performed by accredited auditors reviewing company documentation, or companies can submit reports from supply chain risk analysis platforms to a BIS online compliance platform.}

The purpose of this audit would be to confirm that companies engaged in transactions subject to the EAR have properly conducted due diligence checks as required by BIS guidance---and that they have not engaged in self-blinding practices. The scope of this requirement would depend on BIS's definition of mandatory due diligence, which should ideally include clear reporting triggers (e.g. any customer purchasing $\geq$X restricted chips within a 12-month period).

To make this mechanism operational, BIS could maintain a roster of accredited auditors and accredited supply chain risk analysis platforms authorized to issue compliance reports. These accredited entities could follow a standardized report template and use a standardized, machine-readable schema. The schema should include the customer's identity and any beneficial ownership; corporate control and affiliations; sites of use and intended end use; results from list-based screening (the Entity List, the Office of Foreign Assets Control lists, Military End-User and End-Use lists); any red flags identified and resolution steps for these; and an overall risk assessment (e.g. low, medium, or high). A vulnerability of this mechanism, however, is that it relies on exporters correctly identifying and flagging when customers cross reporting thresholds, creating an incentive for companies motivated to avoid scrutiny to simply not report them. The automated completeness checks mentioned below can mitigate this.

Implementation steps:

\begin{enumerate}
\def\labelenumi{\arabic{enumi}.}
\item
  \textbf{KYC Attestation Form Submission via BIS Online Compliance Platform:} For each applicable customer, the exporter uploads a KYC report generated by an accredited auditor or platform. Each submission is tagged with: customer ID, facility addresses, relevant license numbers, and applicable threshold (chips or compute).
\end{enumerate}

\emph{Performed by: exporters and accredited auditors.}

\begin{enumerate}
\def\labelenumi{\arabic{enumi}.}
\setcounter{enumi}{1}
\item
  \textbf{Automated Completeness Checks:} The platform reconciles submissions against: Export licenses and SNAP-R filings (who you're allowed to sell to); shipment records (EEI/AES ITN, invoices, serial counts). The platform then issues reports for human review, flagging missing KYC for any customer that crosses a threshold and duplicates/mismatches (e.g. report filed for the wrong facility).
\end{enumerate}

\emph{Performed by: BIS.}

\begin{enumerate}
\def\labelenumi{\arabic{enumi}.}
\setcounter{enumi}{2}
\item
  \textbf{Human Review and Sampling:} BIS compliance officers review all high-risk submissions and decide if further action is needed.
\end{enumerate}

\emph{Performed by: BIS.}

\section{End-Use Verification}\label{end-use-verification}

This section examines how to verify the end use of restricted chips across two distinct contexts, each with different actors and verification capabilities.

The first context is chip exporters and server manufacturers selling directly to end customers. Here, the primary concern is preventing restricted actors---such as entities involved in bioweapons development or military-intelligence applications---from acquiring controlled chips. In this context, end-use verification is primarily a pre-sale screening function, and the main tool available is end-use and line-of-business cross-checks, which verify that a customer's declared use is consistent with their actual business activities.

A second scenario---not yet directly relevant but worth flagging for the future---concerns cloud access restrictions under the EAR. As of July 2026, cloud access is not directly regulated under the EAR. BIS does have the authority to impose cloud restrictions as a license condition for AI chip exports, and/or through U.S. persons controls, and there have also been efforts to expand and clarify BIS's authority related to cloud controls.\footnote{U.S. Department of Commerce, ``\href{https://www.bis.gov/media/documents/cloud-based-storefronts-redacted.pdf}{Advisory Opinion on Cloud-based Storefronts.}''} In particular, the Remote Access Security Act, a bill currently under consideration in the U.S. Congress, would empower BIS to regulate remote access to advanced compute when U.S. national security or foreign policy risks are involved.\footnote{Lawler, ``\href{https://www.congress.gov/bill/119th-congress/house-bill/2683}{H.R.2683 - Remote Access Security Act.}''; McCormick, ``\href{https://www.congress.gov/bill/119th-congress/senate-bill/3519}{S.3519 - Remote Access Security Act.}''} If cloud access were indeed brought under the EAR, it could entail certain end-use restrictions, such as prohibitions on the development of WMD, which would require verifying how compute resources are used. In practice, some of these controls already exist: BIS has determined that U.S. persons providing services to train AI models may face a license requirement when there is knowledge that such activities will support WMD or military-intelligence purposes for or on behalf of entities headquartered in a ``D:5 country'' and therefore subject to a U.S. arms embargo under the EAR.\footnote{Bureau of Industry and Security, ``\href{https://www.bis.gov/media/documents/ai-policy-statement-training-ai-models-may-13-2025}{BIS Policy Statement on Controls that May Apply to Advanced Computing Integrated Circuits and Other Commodities Used to Train AI Models.}''} End-use verification would help ensure compliance with these existing rules and any future extensions of them. These mechanisms could also support initiatives such as G42's assurance framework under development for the Pax Silica ecosystem, which aims to bring transparency to compute end-use.\footnote{G42, ``\href{https://www.g42.ai/resources/news/g42-announces-assurance-compute-framework-secure-advanced-us-ai-infrastructure-across-pax-silica-ecosystem}{G42 Announces Assurance Compute Framework to Secure Advanced U.S. AI Infrastructure Across the Pax Silica Ecosystem.}''}

Additionally, the amount of compute that cloud providers supply to customers could also be important to verify in the future. If BIS were to establish restrictions on how much compute foreign cloud providers can sell to non-U.S. customers as a condition of acquiring U.S. chips, verifying compliance with those limits would itself constitute a form of end-use verification.\footnote{An important limitation that applies to both contexts is that the compute required for some of the most concerning end uses (e.g. WMD development) may not be exceptionally large, which means that certain actors may be able to access sufficient compute domestically or through smuggling. End-use verification should therefore be understood as one layer of a broader enforcement strategy rather than a primary safeguard against the most determined adversaries.}

\subsection{End-Use and Line-of-Business Cross-Checks}\label{end-use-and-line-of-business-cross-checks}

\emph{\textbf{Summary:} Inspect records confirming that a customer's line of business aligns with their declared use of compute. This process can be performed by accredited auditors.}

End-use verification would be needed to ensure that advanced compute resources are not used to train AI models for harmful purposes, such as designing DNA sequences for bioweapon development or military-intelligence applications. A non-invasive approach to verifying this would involve cross-checking customers' declared end uses against publicly available business information---for example, unusually high compute use by a biological research lab could raise a red flag. While this method is not immune to circumvention as shell companies can be created to obscure harmful activities, it could still provide a meaningful layer of end-use verification.

To limit the burden on exporters and/or cloud providers, this mechanism could be scoped to customers consuming compute above a defined threshold---for instance, chip orders above a certain level, or customers consuming more than a certain amount of GPU-hours of advanced compute in a quarter. Such a threshold should be periodically reassessed, or fixed to a known variable such as the amount of compute produced annually, to avoid the scope becoming overly broad.

That said, this mechanism would likely be ineffective against sophisticated actors who could establish or acquire front companies whose declared line of business would justify high compute consumption. Considering its limitations, this mechanism is best suited for customers in jurisdictions not considered at high risk of chip diversion, or as one of several verification mechanisms.

Implementation steps:

\begin{enumerate}
\def\labelenumi{\arabic{enumi}.}
\item
  \textbf{Customer Declaration of Use Collection:} Require customers to submit an end-use declaration specifying their line of business, intended workloads, deployment regions, and an acknowledgment of prohibited uses.
\end{enumerate}

\emph{Performed by: exporters and/or cloud providers.}

\begin{enumerate}
\def\labelenumi{\arabic{enumi}.}
\setcounter{enumi}{1}
\item
  \textbf{Business Reality Check:} Verify that the customer's declared end use and line of business are consistent with official records. Such records may include business registry filings from the jurisdiction in which the customer is incorporated and, where available, annual reports, financial statements, and other public documents. Because these records can be difficult to obtain in some jurisdictions and can be falsified if submitted by customers themselves, this task would likely be best performed by accredited auditors with local expertise.
\end{enumerate}

\emph{Performed by: an accredited auditor.}

\begin{enumerate}
\def\labelenumi{\arabic{enumi}.}
\setcounter{enumi}{2}
\item
  \textbf{Submission via BIS Online Compliance Platform:} Once the review is complete, the provider would submit a standardized end-use attestation form confirming that the customer's declared end use is consistent with their verified line of business. The form and supporting materials would then be uploaded to a BIS online compliance platform for automated review.
\end{enumerate}

\emph{Performed by: an accredited auditor.}

\subsection{Compute Provisioning Audits}\label{compute-provisioning-audits}

\emph{\textbf{Summary:} Verify the cumulative compute provisioned by cloud providers to each customer, measured in floating-point operations (FLOP), by reviewing cluster allocation logs, billing records, and usage metrics. These audits could serve as an end-use check, confirming compliance with potential restrictions on how much compute cloud providers can provision to certain restricted customers.}

Compute provisioning audits would verify the cumulative compute (measured in FLOP) that cloud providers allocate to each customer or entity by reviewing cluster allocation logs, billing records, and usage metrics. These audits could function as a type of end-use check in scenarios where BIS establishes compute-based restrictions on foreign cloud providers acquiring U.S. chips---for instance, a condition prohibiting them from provisioning more than a certain amount of FLOP to non-U.S. customers. Given that this mechanism is relatively invasive, it could be scoped to customers consuming compute above a defined threshold---for instance, more than Y GPU-hours in a quarter---in jurisdictions at a high risk of engaging in restricted end uses.

To be effective, this verification method would likely need to be supported by an information-sharing platform among cloud providers, since entities seeking to evade compute restrictions could distribute workloads across multiple providers simultaneously. This system could resemble the information-sharing mechanisms used in the financial sector to detect and prevent fraud, in which institutions share key indicators to identify suspicious patterns across otherwise separate networks.\footnote{Financial Crimes Enforcement Network, ``\href{https://www.fincen.gov/resources/section-314b}{Information Sharing Under Section 314(b).}''}

Implementation steps:\footnote{See Section 3.3.3. Compute Accounting of Heim et al., ``\href{https://www.oxfordmartin.ox.ac.uk/publications/governing-through-the-cloud-the-intermediary-role-of-compute-providers-in-ai-regulation}{Governing Through the Cloud: The Intermediary Role of Compute Providers in AI Regulation.}''}

\begin{enumerate}
\def\labelenumi{\arabic{enumi}.}
\item
  \textbf{Data Collection:} The cloud provider compiles cluster allocation logs and billing or usage records for the audit period, linked to each customer's unique identifier. Where available, this includes telemetry such as utilization or power draw.
\end{enumerate}

\emph{Performed by: cloud provider.}

\begin{enumerate}
\def\labelenumi{\arabic{enumi}.}
\setcounter{enumi}{1}
\item
  \textbf{Compute Usage Accounting:} For each customer, the cloud provider calculates \emph{Compute (FLOP) = Peak FLOP per Second × Utilization × Active Seconds × Number of Accelerators.}\footnote{Floating-point operations (FLOP) is a cumulative measure of computational work---the total number of arithmetic operations performed. Computational power is a rate---the number of operations performed per second (FLOP/s). The two are related by FLOP = FLOP/s × time. For end-use verification, these quantities answer different questions: compute measures how much computational work a customer has done over a period (relevant to training-run thresholds), while computational power measures the capacity available to them at a given moment (relevant to provisioning caps).} If utilization cannot be measured directly, FLOP can be estimated from billed accelerator-hours combined with a default utilization assumption, or inferred from power consumption metrics. \emph{Performed by: cloud provider.}
\item
  \textbf{Aggregation:} The provider sums all jobs per customer for the reporting period and cross-checks against total billed accelerator-hours to ensure consistency. \emph{Performed by: cloud provider.}
\item
  \textbf{Report Submission to the BIS Online Compliance Platform:} Providers submit compute usage reports to BIS or accredited auditors twice a year, listing total compute used (FLOP) per customer, supported by billing and utilization records.
\end{enumerate}

\emph{Performed by: cloud provider.}

\clearpage
\section*{Conclusion}
\phantomsection\label{conclusion}
\addcontentsline{toc}{section}{Conclusion}

The effectiveness of U.S. export controls on AI chips is currently undermined by inefficient enforcement mechanisms and insufficient enforcement resources. The verification systems described in this paper can help address these two challenges in the near term because they can be quickly implemented and involve private-sector actors collaborating with BIS. We recognize that most of the mechanisms proposed here rely, to some degree, on trusting the exporter or chip owner, since they involve validating documents or materials provided by these actors, which could, in principle, be falsified. But while no individual mechanism is impossible to evade, each imposes some costs and friction on bad actors, and these compound when mechanisms are used together.

\section*{Acknowledgements}
\phantomsection\label{acknowledgements}
\addcontentsline{toc}{section}{Acknowledgements}

We are grateful to the following people for discussion and input: Onni Aarne, Mauricio Baker, Will Hodgkins, Arjun Maheshwari, Konstantin Pilz, and Maxwell Roberts. Mistakes and opinions are our own.

\section*{Appendix: Methodology}
\phantomsection\label{appendix-methodology}
\addcontentsline{toc}{section}{Appendix: Methodology}

\textbf{Identifying Candidate Mechanisms:} The authors of this paper identified candidate mechanisms in three ways: by drawing on verification approaches from other domains, such as International Atomic Energy Agency (IAEA) safeguards and financial industry verification practices; by drawing on studies of verification mechanisms proposed for AI governance; and by considering new mechanisms designed to directly address gaps in export control enforcement.

\textbf{Selection Criteria:} The authors selected mechanisms that could be implemented within approximately one year using existing legal authorities, technologies, and/or institutional infrastructure; offer meaningful enforcement value, defined here as a reasonable likelihood of detecting or deterring export control violations; and be cost-effective in the broad sense described below.

\textbf{Cost-Effectiveness:} ``Cost'' in this paper encompasses the following dimensions, which were weighed together:

\begin{itemize}
\item
  \emph{Direct Cost}: The monetary cost of implementing and operating the mechanism, including infrastructure, personnel, and technology.
\item
  \emph{Burden on Private-Sector Actors}: The compliance burden imposed on exporters, cloud providers, and other private entities, including the time and resources required to implement new procedures.
\item
  \emph{Implementation Complexity for BIS}: The degree to which a mechanism requires new regulatory infrastructure, trained personnel, or coordination with other agencies or foreign governments.
\end{itemize}

These cost dimensions were assessed qualitatively rather than through formal modeling.

\textbf{Summary Table Ratings}

\textbf{Maturity Ratings:} The maturity ratings reflect the extent to which each mechanism (or a close analog) has already been implemented in practice, either in the context of export control enforcement or in other verification systems.

\begin{itemize}
\item
  Established: The mechanism is already in use, either in the context of export control enforcement, other verification systems, or both.
\item
  Relatively Established: The mechanism is similar to one that is already established but would require some modification to be applied as proposed in this paper.
\item
  Novel: The mechanism is not currently implemented in the context of export control enforcement or other verification systems.
\end{itemize}

\textbf{Effectiveness Ratings:} The effectiveness ratings consider a threat model comprised of three types of adversaries: (1) well-resourced, state-linked entities seeking to acquire or use restricted compute; (2) commercial actors seeking to acquire chips through shell companies or intermediaries to circumvent export restrictions; and (3) companies set up primarily to broker sales to unauthorized destinations, serving as intermediaries between suppliers and restricted users.

The ratings reflect how reliably each mechanism would detect a violation by any of these adversaries:

\begin{itemize}
\item
  High: The mechanism is difficult to evade across the range of adversaries described above.
\item
  Medium: The mechanism provides meaningful assurance but has evasion paths accessible to a sophisticated actor.
\item
  Low: The mechanism is useful for flagging issues but limited as a stand-alone enforcement tool, as it can be defeated by a moderately resourced adversary acting alone.
\end{itemize}

These ratings are necessarily qualitative and reflect the judgment of the authors rather than empirical testing.

\textbf{Degree of Access Required:} This rating reflects the extent to which a mechanism requires access to a regulated entity's physical premises, internal systems, or proprietary and sensitive information.

\begin{itemize}
\item
  Invasive: The mechanism requires either direct physical access to an entity's premises or access to sensitive, proprietary information about an entity's operations.
\item
  Relatively Invasive: The mechanism requires limited or remote access to an entity's facilities or systems, without ongoing access to proprietary operational data (e.g. remote video walkthroughs).
\item
  Not Invasive: The mechanism relies on information that is publicly available or already reported by the entity, without requiring additional access to facilities, systems, or proprietary data.
\end{itemize}

\section*{Bibliography}
\phantomsection\label{bibliography}
\addcontentsline{toc}{section}{Bibliography}
\small
\setlength{\parskip}{5pt}

\bibentrystart Aarne, Onni and Erich Grunewald. ``Export Auditors as Market-Powered Export Enforcement.'' Institute for AI Policy and Strategy, March 2026. \url{https://static1.squarespace.com/static/64edf8e7f2b10d716b5ba0e1/t/69bc642c9cbbd038d27a9660/1773954092111/Export+Auditors+as+Market-Powered+Export+Enforcement.pdf}.

\bibentrystart AEB. ``Straightforward and Secure Restricted Party Screening.'' April 2025. \url{https://www.aeb.com/en/compliance-screening/index.php}.

\bibentrystart Axis Communications. ``Product Support for AXIS Q6078-E PTZ Camera.'' January 2023. \url{https://www.axis.com/products/axis-q6078-e/support}.

\bibentrystart Axis Communications. ``AXIS Q6078-E PTZ Camera.'' Accessed August 16, 2026. \url{https://www.axis.com/dam/public/22/dc/ba/datasheet-axis-q6078-e-ptz-camera-en-US-487394.pdf}.

\bibentrystart Baker, Mauricio, Gabriel Kulp, Oliver Marks, Miles Brundage, and Lennart Heim. ``Verifying International Agreements on AI: Six Layers of Verification for Rules on Large-Scale AI Development and Deployment.'' July 2025. \url{https://arxiv.org/abs/2507.15916}.

\bibentrystart Baute, Jacques. ``Timeline IRAQ: Challenges \& Lessons Learned from Nuclear Inspections.'' June 2004. \url{https://www.iaea.org/sites/default/files/publications/magazines/bulletin/bull46-1/46102486468.pdf}.

\bibentrystart Brass, Asher and Onni Aarne. ``Location Verification for AI Chips.'' Institute for AI Policy and Strategy, April 2024. \url{https://static1.squarespace.com/static/64edf8e7f2b10d716b5ba0e1/t/6670467ebe2a477eb1554f40/1718634112482/Location%2BVerification%2Bfor%2BAI%2BChips.pdf}.

\bibentrystart Bureau of Industry and Security. ``BIS Policy Statement on Controls That May Apply to Advanced Computing Integrated Circuits and Other Commodities Used to Train AI Models.'' May 2025. \url{https://www.bis.gov/media/documents/ai-policy-statement-training-ai-models-may-13-2025}.

\bibentrystart Code of Federal Regulations. ``Part 744---Control Policy: End-User and End-Use Based.'' March 2017. \url{https://www.ecfr.gov/current/title-15/subtitle-B/chapter-VII/subchapter-C/part-744#Supplement-No.-7-to-Part-744}.

\bibentrystart Code of Federal Regulations. ``§ 758.7 Authorities of the Bureau of Industry and Security, Office of Export Enforcement (OEE).'' November 2020. \url{https://www.ecfr.gov/current/title-15/subtitle-B/chapter-VII/subchapter-C/part-758/section-758.7}.

\bibentrystart ———. ``§ 764.2 Violations.'' November 2020. \url{https://www.ecfr.gov/current/title-15/subtitle-B/chapter-VII/subchapter-C/part-764/section-764.2}.

\bibentrystart ———. ``Supplement No. 3 to Part 732---BIS's ``Know Your Customer'' Guidance and Red Flags.'' October 2023. \url{https://www.ecfr.gov/current/title-15/subtitle-B/chapter-VII/subchapter-C/part-732/appendix-Supplement%20No.%203%20to%20Part%20732}.

\bibentrystart ———. ``§ 743.9 Reporting Requirements for ``Front-End Fabricators'' Producing ``Applicable Advanced Logic Integrated Circuits'' for Authorized Integrated Circuit Designers.'' January 2025. \url{https://www.ecfr.gov/current/title-15/subtitle-B/chapter-VII/subchapter-C/part-743/section-743.9}.

\bibentrystart Cui, Shengkun, Archit Patke, Hung Nguyen, et al. ``Story of Two GPUs: Characterizing the Resilience of Hopper H100 and Ampere A100 GPUs.'' arXiv, March 2025. \url{https://arxiv.org/abs/2503.11901}.

\bibentrystart Financial Crimes Enforcement Network. ``Information Sharing Under Section 314(b).'' September 2012. \url{https://www.fincen.gov/resources/section-314b}.

\bibentrystart Fist, Tim and Erich Grunewald. ``Preventing AI Chip Smuggling to China.'' Center for a New American Security, October 2023. \url{https://www.cnas.org/publications/reports/preventing-ai-chip-smuggling-to-china}.

\bibentrystart Fournier, Vincent. ``Surveying Safeguarded Material 24/7.'' International Atomic Energy Agency, September 2016. \url{https://www.iaea.org/newscenter/news/surveying-safeguarded-material-24/7}.

\bibentrystart G42. ``G42 Announces Assurance Compute Framework to Secure Advanced U.S. AI Infrastructure Across the Pax Silica Ecosystem.'' February 2026. \url{https://www.g42.ai/resources/news/g42-announces-assurance-compute-framework-secure-advanced-us-ai-infrastructure-across-pax-silica-ecosystem}.

\bibentrystart Grunewald, Erich. ``How Banned AI Chips End up in China.'' \emph{The Substrate}, May 2026. \url{https://www.the-substrate.net/p/how-banned-ai-chips-end-up-in-china}.

\bibentrystart Heim, Lennart, Tim Fist, Janet Egan, et al. ``Governing Through the Cloud: The Intermediary Role of Compute Providers in AI Regulation.'' March 2024. \url{https://www.oxfordmartin.ox.ac.uk/publications/governing-through-the-cloud-the-intermediary-role-of-compute-providers-in-ai-regulation}.

\bibentrystart Huizenga, Bill. ``H.R.3447 - Chip Security Act.'' \emph{Congress.Gov}, May 2025. \url{https://www.congress.gov/bill/119th-congress/house-bill/3447/text}.

\bibentrystart Industry and Security Bureau. ``Expansion of Validated End User Authorization: Data Center Validated End User Authorization.'' \emph{Federal Register: The Daily Journal of the United States Government}, October 2024. \url{https://www.federalregister.gov/documents/2024/10/02/2024-22587/expansion-of-validated-end-user-authorization-data-center-validated-end-user-authorization}.

\bibentrystart International Atomic Energy Agency. ``Equipment Used in Safeguards.'' November 2010. \url{https://www.iaea.org/newscenter/multimedia/photoessays/equipment-used-safeguards}.

\bibentrystart International Trade Administration. ``Common Export Documents.'' January 2020. \url{https://www.trade.gov/common-export-documents}.

\bibentrystart Lawler, Michael. ``H.R.2683 - Remote Access Security Act.'' \emph{Congress.Gov}, April 2025. \url{https://www.congress.gov/bill/119th-congress/house-bill/2683}.

\bibentrystart McCormick, David. ``S.3519 - Remote Access Security Act.'' \emph{Congress.Gov}, December 2025. \url{https://www.congress.gov/bill/119th-congress/senate-bill/3519}.

\bibentrystart Nellis, Stephen and Michael Martina. ``Nvidia Builds Location Verification Tech That Could Help Fight Chip Smuggling.'' \emph{Reuters}, December 2025. \url{https://www.reuters.com/business/nvidia-builds-location-verification-tech-that-could-help-fight-chip-smuggling-2025-12-10/}.

\bibentrystart NVIDIA Newsroom. ``Opt-In NVIDIA Software Enables Data Center Fleet Management.'' \emph{NVIDIA}, December 2025. \url{https://blogs.nvidia.com/blog/optional-data-center-fleet-management-software/}.

\bibentrystart Roberts, Maxwell K. ``BIS Is Getting More Funding---Here's How to Spend It.'' \emph{The Substrate}, January 2026. \url{https://www.the-substrate.net/p/bis-is-getting-more-fundingheres}.

\bibentrystart ———. ``BIS Should Build a Lean, Mean, Data-Driven Enforcement Machine.'' \emph{The Substrate}, February 2026. \url{https://www.the-substrate.net/p/bis-should-build-a-lean-mean-data}.

\bibentrystart Select Committee on China. ``House Committee Passes Chip Security Act.'' March 2026. \url{https://chinaselectcommittee.house.gov/media/press-releases/house-committee-passes-chip-security-act}.

\bibentrystart Sequeira, Vitor, Erik Wolfart, Gunnar Boström, et al. ``Laser Curtain for Containment and Tracking.'' September 2021. \url{https://resources.inmm.org/sites/default/files/2021-09/a498.pdf}.

\bibentrystart SGS. ``ISCC PLUS Certification.'' May 2023. \url{https://www.sgs.com/en/services/iscc-plus-certification}.

\bibentrystart ———. ``EU Deforestation Regulation (EUDR) Chain of Custody Verification.'' July 2026. \url{https://www.sgs.com/en/services/eu-deforestation-regulation-eudr-chain-of-custody-verification}.

\bibentrystart ``SolidIntel.'' Accessed August 16, 2026. \url{https://solidintel.com/}.

\bibentrystart United Nations. ``Sixth Report of the Executive Chairman of the Special Commission.'' December 1993. \url{https://www.un.org/Depts/unscom/sres26910.htm}.

\bibentrystart United States District Court Southern District Of New York. ``UNITED STATES OF AMERICA V. YIH-SHYAN `Wally' LIAW, RUEI-TSANG `Steven' CHANG, and TING-WEI `Willy' SUN.'' March 2026. \url{https://storage.courtlistener.com/recap/gov.uscourts.nysd.660030/gov.uscourts.nysd.660030.2.0.pdf}.

\bibentrystart U.S. Department of Commerce. ``Advisory Opinion on Cloud-Based Storefronts.'' November 2014. \url{https://www.bis.gov/media/documents/cloud-based-storefronts-redacted.pdf}.

\bibentrystart ———. ``Denied Persons List (DPL).'' March 2025. \url{https://www.bis.gov/licensing/end-user-guidance/denied-persons-list-dpl}.

\bibentrystart ———. ``Part 744 - Control Policy: End-User and End-Use Based.'' October 2025. \url{https://www.bis.gov/regulations/ear/744#section-744.16}.

\bibentrystart ———. ``Penalties.'' August 2025. \url{https://www.bis.gov/enforcement/penalties}.

\bibentrystart ———. ``Supplement No. 3 to Part 732---BIS's ``Know Your Customer'' Guidance and Red Flags.'' March 2025. \url{https://www.bis.gov/node/1533}.

\bibentrystart ———. ``Fiscal Year 2027 President's Budget Request.'' April 2026. \url{https://www.commerce.gov/sites/default/files/2026-04/FY2027-BIS-CJ-Submission.pdf}.

\bibentrystart U.S. Department of Commerce, U.S. Department of Treasury, and U.S. Department of Justice. ``Obligations of Foreign-Based Persons to Comply with U.S. Sanctions and Export Control Laws.'' March 2024. \url{https://www.justice.gov/archives/opa/media/1341411/dl?inline}.

\bibentrystart U.S. Department of Treasury. ``Sanctions List Search Tool.'' Accessed August 16, 2026. \url{https://ofac.treasury.gov/sanctions-list-search-tool}.

\bibentrystart ``WireScreen.'' Accessed August 16, 2026. \url{https://wirescreen.ai/}.

\end{document}